\documentclass[aps,pre,onecolumn,superscriptaddress,notitlepage,longbibliography]{revtex4-1}

\usepackage[pdftex]{hyperref}
\usepackage{graphicx}
\usepackage[caption=false]{subfig}
\usepackage{amssymb}
\usepackage{amsmath}
\usepackage{verbatim}
\usepackage{bm}
\usepackage{bbold}
\usepackage[separate-uncertainty=true,multi-part-units=single]{siunitx}
\usepackage{color}
\usepackage[american]{babel}
\DeclareSIUnit\micron{\micro\metre}
\usepackage{physics}
\usepackage{mathtools}

\newcommand{\del}{\partial}
\newcommand{\RRR}{\mathcal{R}}

\newcommand{\al}[1]{\begin{align}#1\end{align}}
\newcommand{\eqn}[1]{\begin{equation}#1\end{equation}}
\newcommand{\ut}[1]{_\text{#1}}
\newcommand{\vect}[1]{\boldsymbol{#1}}
\newcommand{\pars}[1]{\left({#1}\right)}
\newcommand{\brcs}[1]{\lbrace{#1}\rbrace}
\newcommand{\avg}[1]{\langle{#1}\rangle}
\newcommand{\pder}[2]{\frac{\partial #1}{\partial #2}}

\begin{document}


\title{Calculating the motion of highly confined, arbitrary-shaped particles in Hele-Shaw channels}


\author{Bram Bet}
\email[]{B.p.bet@uu.nl}
\affiliation{Institute for Theoretical Physics, Center for Extreme Matter and Emergent Phenomena, Utrecht University, Princetonplein 5, 3584 CC Utrecht, The Netherlands}

\author{Rumen Georgiev}%
\affiliation{Process and Energy Department, Delft University of Technology, Delft 2628 CD, The Netherlands}

\author{William Uspal}
\affiliation{Max Planck Institute for Intelligent Systems, Heisenbergstr. 3, 70569 Stuttgart, Germany}
\affiliation{Institut f\"ur Theoretische Physik, Universität Stuttgart, Pfaffenwaldring 57, D-70569 Stuttgart, Germany}

\author{Huyesin Burak Eral}%
\affiliation{Process and Energy Department, Delft University of Technology, Delft 2628 CD, The Netherlands}

\author{Ren\'e van Roij}%
\affiliation{Institute for Theoretical Physics, Center for Extreme Matter and Emergent Phenomena, Utrecht University, Princetonplein 5, 3584 CC Utrecht, The Netherlands}

\author{Sela Samin}%
\affiliation{Institute for Theoretical Physics, Center for Extreme Matter and Emergent Phenomena, Utrecht University, Princetonplein 5, 3584 CC Utrecht, The Netherlands}


\date{\today}


\begin{abstract}
We combine theory, numerical calculations, and experiments to 
accurately predict the motion of anisotropic particles in shallow microfluidic 
channels, in which the particles are strongly confined in the vertical 
direction. We formulate an effective quasi-two-dimensional description of the 
Stokes flow around the particle via the Brinkman equation, which can be solved 
in a time that is two orders of magnitude faster than the three-dimensional 
problem. The computational speedup enables us to calculate the full trajectories 
of particles in the channel. To test our scheme, we study the 
motion of dumbbell-shaped particles that are produced in a microfluidic 
channel using `continuous flow lithography'. Contrary to what was reported in 
earlier work (\citeauthor{uspal2013engineering}, Nature communications {\bf 4} (2013)), we 
find that the reorientation time of a dumbbell particle in an external flow 
exhibits a minimum as a function of its disk size ratio. This finding is in 
excellent agreement with new experiments, thus confirming the predictive power 
of our scheme.
\end{abstract}

\pacs{}

\maketitle

\section{Introduction}
Microfluidic devices offer many applications, such as flow cytometry \cite{oakey2010particle, wang2007dielectrophoresis,mao2009single}, separation of cells \cite{gossett2010label}, DNA sequencing \cite{tewhey2009microdroplet}, or blood cell analysis \cite{toner2005blood}. In many applications, it is of paramount importance to control the position of the immersed particles. Often, focusing of particles in the channel is achieved by external fields or by flows induced by the channel geometry \cite{xuan2010particle}, while particle separation or sorting can also be realised by tuning electric fields \cite{jeon2016continuous,wang2007dielectrophoresis} or the channel geometry \cite{gossett2010label,sajeesh2014particle,pamme2007continuous,zeming2013rotational,li2014inertial}. Alternatively, the shape of the particle itself offers a different route to manoeuvring the particles in the channel \cite{masaeli2012continuous}. The hydrodynamics of various particles in strong confinement 
has been extensively studied, for instance the motion of confined droplets \cite{beatus2012physics,shen2014dynamics}, (connected) disks \cite{uspal2012scattering,uspal2013engineering}, and fibers \cite{berthet2013single}. However, as tunability increases with the complexity of the particle shapes, theoretical arguments on the basis of simplified geometries might fall short of accurately describing the motion of particles. In addition, advances in versatile particle synthesis techniques, such as continuous-flow lithography \cite{dendukuri2006continuous}, make an infinite variety of quasi-two-dimensional shapes experimentally accessible. 
\begin{figure}
\subfloat[Top view.\label{fig:geomltop}]{%
  \includegraphics[width=0.6\textwidth]{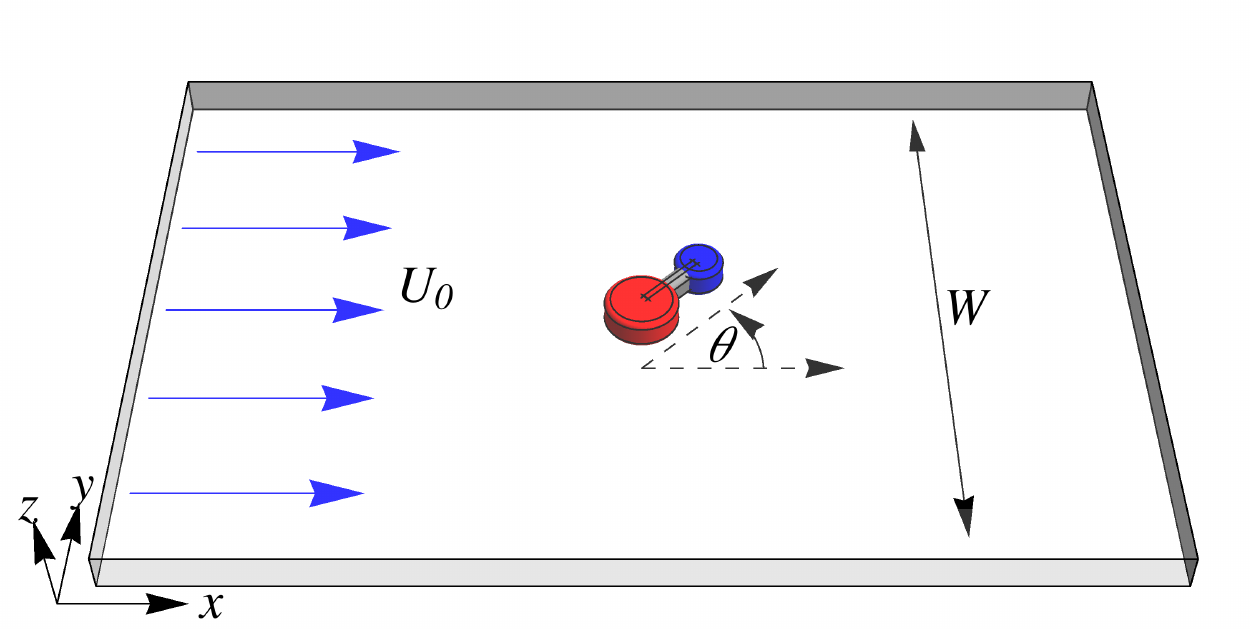}%
} \\
\subfloat[Side view.\label{fig:geomltop}]{%
  \includegraphics[width=0.6\textwidth]{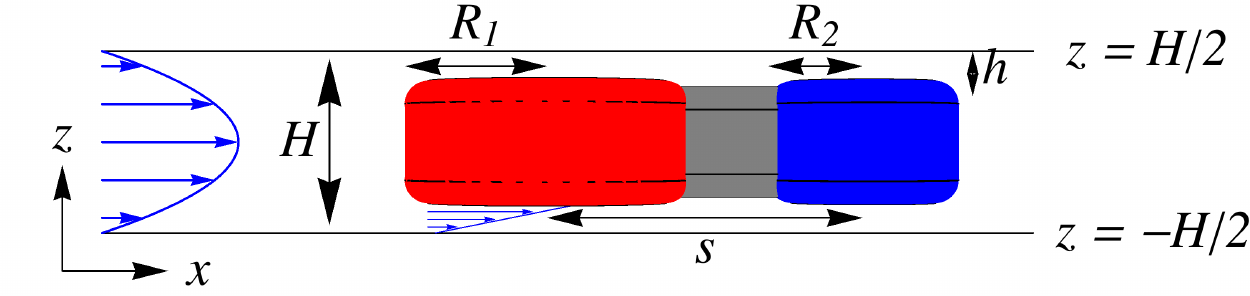}%
}
\caption{Top view (a) and side view (b) of the geometry and the Cartesian frame. An external flow $\bm{U}_0$ with parabolic profile is imposed through a channel of length $L$, width $W$ and height $H$, containing a dumbbell particle with radii $R_1$ and $R_2$ at a center-to-center distance $s$, with height $H-2 h$, such that the height of the gaps between the particle and the top and bottom walls is given by $h$. In these gaps, the flow profile is approximately that of a simple shear flow. The orientation of the long axes of the dumbbell with respect to the external flow $\bm{U}_0$ is denoted by $\theta$.} \label{fig:geom}
\end{figure}

Inspired by these advances, we develop in this work a method to calculate the two-dimensional motion of confined particles in microfluidic channels that can handle particles of any given quasi-two-dimensional shape. Here, we combine finite-element calculations with a simple approach for the particle motion to solve the full hydrodynamic equations at hand, either in full detail in the three-dimensional geometry or in an effective two-dimensional description. Our method is validated by comparison of the three- and two-dimensional results with analytical calculations. Next, we apply our method to study dumbbell-shaped particles in Hele-Shaw channels, reproducing 
new experimental results accurately, without any adjustable parameters.

\section{Theory \& Numerical Methods}
\subsection{Hydrodynamic equations and equations of motion} \label{sec:stokeseom}
We consider a rigid particle, of arbitrary shape, at 
position $\bm{r}_p = (x_p,y_p,z_p)$, which is immersed in a fluid 
that is driven through a shallow microfluidic channel by the application of 
pressure at the channel inlet, see the illustration in Fig. \ref{fig:geom}. 
Let us assume that this particle is strongly confined in the 
$z$-direction, i.e. the particle is separated from the top and bottom wall by a 
small gap of height $h$ that is much smaller than the channel height $H$. Such a 
particle can be produced, for example, using `continuous flow lithography' 
\cite{dendukuri2006continuous}. In this method, the fluid in the channel 
is a pre-polymer solution, where particles are `printed' by cross-linking the 
oligomers by pulses of UV light, which is applied through a photolithographic 
mask. In this way, the particle shape is defined in the $xy$-plane by the shape 
of the mask (which can be of any desired shape), while it is extruded in the 
$z$-direction to a height that is comparable to the channel height, such that $h 
\ll H$.
%

The microfluidic channel under consideration here (see Appendix \ref{app:exp}) 
has a height $H = \SI{30}{\micro \meter}$ and a much larger width $W = 
\SI{500}{\micro \meter}$, while the length $L$ of the channel is of the order of one 
centimeter, which can be considered infinite for our analysis. The fluid is 
driven through the channel at an average velocity $U_0$, which is of the order 
of $\SI{50}{\micro \meter \per \second}$. Using the hydraulic diameter 
$D_H=2HW/(H+W)$ as the characteristic length scale, we find that the Reynolds 
number is $\si{Re}=\rho U_0 D_H/\eta = \SI{e-4}{} - \SI{e-5}{}$ for a typical 
oligomer solution with viscosity $\eta$ and density $\rho$. Therefore, 
the flow is well described by the Stokes equation
\cite{kim1991,happel2012low,leal2007advanced}:
\eqn{ \label{eq:stokes}
 - \nabla p + \eta \nabla^2 \bm{u} =0,  \qquad \nabla \cdot \bm{u} =0,	
}
where $\bm{u}(\bm{r})$ and $p(\bm{r})$ are the fluid velocity and pressure at position $\bm{r}$, respectively, and $\eta$ denotes the fluid viscosity. We supplement the Stokes equation with no-slip boundary conditions on the (stationary) channel walls and the (moving) particle surface $S_p$,
\eqn{ \label{eq:wall_bc}
\bm{u}\big|_{\text{walls}} = 0, \quad \bm{u}\big|_{\bm{r} \in S_p} = \bm{U}_{p} +  \bm{\omega}_p \times (\bm{r}-\bm{r}_p),
}
where $\bm{U}_{p} $ and $\bm{\omega}_p$ denote the particle velocity and angular velocity, respectively. At the inlet of the channel, we impose a uniform incoming Hele-Shaw flow
\eqn{ \label{eq:inlet_bc}
 \bm{u}\big|_{\text{inlet}} = \bm{U}_0(\bm{r}) = \frac{3}{2}\left(1-\frac{4 z^2}{H^2}\right) (U_0,0,0),
 }
which is the analytic solution for the flow between two infinite parallel plates driven by a constant pressure difference. Here, $z \in [-H/2,H/2]$ and $U_0$ is the average velocity in the $x$-direction. Note that this boundary condition neglects the no-slip condition on the sidewalls, i.e., it is not accurate close to the side walls. Therefore, we take into account a finite `entrance length' at the inlet after which the flow is fully developed downstream in the channel. Finally, we impose a zero pressure boundary condition at the outlet, such that the pressure difference between inlet and outlet is precisely driving the flow field given by Eq. \eqref{eq:inlet_bc}, i.e., $U_0 = - H^2 \nabla p /(12 \eta)$. The influence of the no-slip side walls on this pressure drop is assumed to be negligible. The Stokes equation \eqref{eq:stokes}, together with these boundary conditions, form a closed set of equations that we solve numerically by a finite-elements scheme (see Section \ref{sec:numericalmethods}). 

The fluid in the channel surrounding the particle exerts a hydrodynamic force $\bm{F}$ and torque $\bm{T}$ on the particle, which is calculated by integrating the hydrodynamic stress tensor, $\sigma_{ij} = -p \delta_{ij} + \eta (\del_i u_j + \del_j u_i)$, over the particle surface:
 \eqn{  \label{eq:hydroforce}
 \bm{F} = \int_{S_p}dS \ \sigma \cdot \bm{n}, \quad \bm{T} = \int_{S_p}dS \ (\bm{r}- \bm{r}_p) \times (\sigma \cdot \bm{n}).
 }
Due to the linearity of the Stokes equation \eqref{eq:stokes}, we can write the solution $(p,\bm{u})$ of Eq. \eqref{eq:stokes} with boundary conditions \eqref{eq:wall_bc} and \eqref{eq:inlet_bc} as a superposition of two solutions \cite{happel2012low}: $\bm{u} = \bm{u}_0 + \bm{u}'$ and $p = p_0 + p'$, where $(p_0,\bm{u}_0)$ and $(p',\bm{u}')$ are solutions to the Stokes equation \eqref{eq:stokes} with boundary conditions
\al{
\bm{u}_0\big|_{\text{walls}} &= 0, \quad &&\bm{u}_0\big|_{\text{inlet}} = \bm{U}_0(\bm{r}), \quad && \bm{u}_0\big|_{\bm{r} \in S_p} = 0; \label{eq:subprob1}\\
\bm{u}'\big|_{\text{walls}} &= 0,  \quad &&\bm{u}'\big|_{\text{inlet}} = 0, \quad && \bm{u}'\big|_{\bm{r} \in S_p} = \bm{U}_{p} + \bm{\omega}_p \times (\bm{r}-\bm{r}_p) \label{eq:subprob2}, 
}
that is, $\bm{u}_0$ is the solution where the particle is fixed subject to the imposed external flow, and $\bm{u}'$ the solution where the particle moves through the channel without an imposed flow.
The stress tensor also splits accordingly, $\sigma = \sigma_0 + \sigma'$, such that we find that the forces and torque on the particle are written as $\bm{F} = \bm{F}_0 + \bm{F}'$ and $\bm{T}= \bm{T}_0 + \bm{T}'$, which are calculated from Eq. \eqref{eq:hydroforce} by replacing $\sigma$ with $\sigma_0$ or $\sigma'$.

To proceed, we can again use the linearity of the Stokes equation \eqref{eq:stokes} to derive that the force $\bm{F}'$ and torque $T'$ depend linearly on each component of the particle (angular) velocity via a $6 \times 6$ resistance tensor $\RRR$ as \cite{happel2012low,brenner1963stokes,brenner1964stokes}
\eqn{  \label{eq:forceprime}
\left(\begin{array}{c}\bm{F}' \\ \bm{T}'\end{array}\right) = - \eta \RRR \left(\begin{array}{c}\bm{U}_p \\ \omega_p\end{array}\right).
}
Due to the over-damped nature of the system, the (hydrodynamic) force and torque on the particle vanish in the absence of external forces on the particle. Therefore, after summing up the force contributions from the solutions $\bm{u}_0$ and $\bm{u}'$, we find that the particle must obey the equation of motion
\eqn{ \label{eq:eom}
\left(\begin{array}{c}\bm{F} \\ \bm{T}\end{array}\right) = - \eta \RRR \left(\begin{array}{c}\bm{U}_p \\ \bm{\omega}_p\end{array}\right) +\left(\begin{array}{c}\bm{F}_0 \\ \bm{T}_0 \end{array}\right) = \left(\begin{array}{c}\bm{0} \\ \bm{0}\end{array}\right), 
}
with $\bm{0} = (0,0,0)$. Thus, once $\RRR, \bm{F}_0$ and $\bm{T}_0$ are determined, either analytically or numerically, the equations of motion \eqref{eq:eom} can be solved for the particle velocity and angular velocity as
\eqn{ \label{eq:eomsol3D}
\left(\begin{array}{c}\bm{U}_p \\ \bm{\omega}_p\end{array}\right) = \frac{1}{\eta} \RRR^{-1} \left(\begin{array}{c}\bm{F}_0 \\ \bm{T}_0 \end{array}\right) .
}
Notice that $\RRR, \bm{F}_0$ and $\bm{T}_0$ depend on the position and orientation of the particle, such that Eq. \eqref{eq:eomsol3D} only determines the force- and torque-free (angular) velocity for that specific particle position and velocity. The position dependence of $\RRR$ is related to the effects of the side walls; in the case of an infinite slit, the tensor $\RRR$ will only depend on the particle geometry and its orientation (with respect to the imposed external flow).

In this work, due to the strong confinement in the vertical direction as well as the mirror symmetry in the $z=0$ plane of the problem at hand, the movement of the particle is restricted to the two-dimensional mid-plane of the channel at $z=0$, thereby reducing the number of degrees of freedom to three: $\bm{U}_{p} = (U_{p,x},U_{p,y},0)$ and $\bm{\omega}_p = (0,0,\omega_p)$, and similarly $\bm{F} = (F_x , F_y ,0)$ and $\bm{T} = (0 , 0, T)$ \footnote{For the remainder of this text, $\bm{F}$ and $\bm{U}_p$ denote two-dimensional force and particle velocity, respectively.}. Moreover, the particle position is determined by the coordinates $(x_p , y_p)$, while its orientation is described by a single angle $\theta$, as illustrated in Fig. \ref{fig:geom}. Finally, the resistance tensor $\RRR$ reduces to a $3 \times 3$ tensor, which is obtained from the relevant components of the original $6 \times 6$ resistance tensor. In some cases, symmetry arguments may be invoked to reduce the number of degrees of freedom even further, e.g. mirror symmetric particles that are aligned with the imposed external flow, as we shall see below.
%
%
 
Using a finite-elements scheme we can numerically solve the flow field $\bm{u}$ for any  imposed velocity and angular velocity, and from these solutions obtain the resistance tensor $\RRR$, the force $\bm{F}_0$ and the torque $T_0$, to subsequently obtain the force- and torque-free velocities from Eq.  \eqref{eq:eomsol3D}. By repeating this at the new particle position in the next time step, it is in principle possible to integrate the complete particle motion.%
 
\subsection{Brinkman equation}
A three-dimensional finite-elements calculation is able to resolve the flow in 
the channel. However, due to a separation of length scales, $h \ll W$, the 
finite-element mesh needs to be chosen very fine at certain places, causing the 
calculations to be computationally costly, and eventually prohibitive for the 
purpose of integrating the particle motion. To circumvent 
this problem, we resort to an effective 2D-description of the system via the Brinkman 
equation \cite{brinkman1949calculation,uspal2014self}. 
 
Far enough from the side walls and the particle, the flow field is well described by Hele-Shaw flow:
\eqn{ \label{eq:height}
\bm{u}(x,y,z) = \frac{3}{2}\bar{\bm{u}} (x,y) \left(1 - \frac{ 4 z^2}{H^2} \right),
}
the pressure being independent of $z$ to a good approximation: $p = p(x,y)$. We substitute this ansatz in the Stokes equation \eqref{eq:stokes} and average over the channel height to find the Brinkman equation \cite{brinkman1949calculation,uspal2014self}
\eqn{ \label{eq:brinkman}
- \nabla_{\text{2D}} \bar{p} + \eta H \nabla_{\text{2D}}^2 \ \bar{\bm{u}} - \frac{12 \eta }{ H} \bar{\bm{u}} =0, \quad \nabla_{\text{2D}} \cdot  \bar{\bm{u}}=0,
}
where the two-dimensional vector field $\bar{\bm{u}} = (\bar{u}_x,\bar{u}_y)$ 
denotes the $z$-averaged value of the three-dimensional flow field $\bm{u}$, 
and $\bar{p} = p H$ denotes the two-dimensional pressure field. Henceforth, we 
will denote the two-dimensional height-averaged quantities with an overbar. The 
boundary conditions that supplement Eq. \eqref{eq:brinkman} are accordingly:
\al{ \label{eq:brinkman_bc}
\bar{\bm{u}}\big|_{\text{walls}} = 0,  \quad \bar{\bm{u}}\big|_{\text{inlet}} = U_0 \hat{\bm{x}}, \quad \bar{\bm{u}}\big|_{(x,y) \in \del \bar{S}_p} = \bm{U}_{p} +   \omega_p (-(y-y_p),x-x_p),
}
where $\bm{r} = (x,y)$ and $\del \bar{S}_p$ denotes the one-dimensional 
particle boundary of the projected particle surface 
$\bar{S}_p$. Notice that the walls in this situation are the projections of the 
sidewalls at $y = \pm W/2$. As before, $\bm{U}_p = (U_{p,x},U_{p,y})$ and 
$\omega_p$ denote the particle velocity and angular velocity, respectively. 

The solution $(\bar{p},\bar{\bm{u}})$ of the Brinkman equation \eqref{eq:brinkman} defines a stress tensor 
\eqn{ \label{eq:brinkmanstress}
\bar{\sigma}_{ij} = -\delta_{ij} \bar{p} + \eta H (\del_i \bar{u}_j + \del_j \bar{u}_i),
} 
which is integrated over the particle surface to find the hydrodynamic force and torque on the particle:
\al{
\bm{F}_f = \int_{\del \bar{S}_p} ds \ \bar{\sigma} \cdot \bm{n}, \quad T_f =  \int_{\del \bar{S}_p} ds \ \bigg(  (x-x_p)(\bar{\sigma} \cdot \bm{n})_y - (y-y_p)(\bar{\sigma} \cdot \bm{n})_x \bigg). \label{eq:brinkmantorque}
}
The subscript `$f$' indicates that this is the force and torque due to the 
surrounding two-dimensional fluid, and \emph{does not} include the force and 
torque from the confining walls.

Similar to the Stokes equation, the Brinkman equation is linear in the fields 
$\bar{p}$ and $\bar{u}$. Using a similar decomposition as in the previous 
section, we prove explicitly in Appendix 
\ref{app:brinkmanresistance} that the force $\bm{F}_f$ and torque $T_f$ admit a 
linear relation to the particle velocity and angular velocity in terms of a $3 
\times 3$ resistance tensor $\RRR_f$:
\eqn{ \label{eq:fluideom}
\left(\begin{array}{c}\bm{F} \\ T\end{array}\right)_f = -\eta \RRR_f \left(\begin{array}{c}\bm{U}_p \\ \omega_p \end{array}\right)  +  \left(\begin{array}{c}\bm{F}_0 \\ T_0 \end{array}\right).
}
In Appendix \ref{app:brinkmanresistance}, it is shown that that $\RRR_f$ is 
symmetric by employing a version of the Lorentz reciprocal theorem for 
solutions of the Brinkman equation, which is given in Appendix 
\ref{app:lorentz}.

Since the fluid-filled gaps between the particle and the walls are not accounted 
in the Brinkman description, their contribution to the force and torque on the 
particle, which stem from the friction with the top and bottom wall, are 
missing. To obtain this contribution, we assume locally a simple shear flow in 
the narrow gaps between the particle and the confining walls. As 
a result, an area element $dS$ on either of the particle faces that moves with 
a velocity $\bm{v}_S = \bm{U}_p + \bm{\omega}_p \times (\bm{r}-\bm{r}_p)$ with 
respect to the wall, will experience a friction force $\bm{f}_S = - (\eta/h) 
\bm{v}_SdS$. The total force $\bm{F}_w$ on the particle due to the wall friction 
is then found by integrating over the particle-gap surface,
\eqn{ \label{eq:wallforce}
\bm{F}_w  = -  \frac{2\eta}{h} \int_{\bar{S}_p} dS \ \left( \bm{U}_p + \bm{\omega}_p \times (\bm{r}-\bm{r}_p) \right),
}
where the factor $2$ is due to the two gaps. The area element $dS$ also generates a torque $\bm{r} \times \bm{f}_S$, which can be integrated to find the frictional torque on the particle due to the walls:
\eqn{ \label{eq:walltorque}
\bm{T}_w = -  \frac{2\eta}{h} \int_{\bar{S}_p} dS \ \bigg((\bm{r}-\bm{r}_p) \times \left( \bm{U}_p + \bm{\omega}_p \times (\bm{r}-\bm{r}_p) \right) \bigg).
}
As before, the linear dependence of Eqs. \eqref{eq:wallforce} and \eqref{eq:walltorque} in the particle velocities leads to
\eqn{ \label{eq:walleom}
\left(\begin{array}{c}\bm{F} \\ T\end{array}\right)_w = -\eta\RRR_w \left(\begin{array}{c}\bm{U}_p \\ \omega_p \end{array}\right),
}
where the components of the $3 \times 3$ wall resistance tensor $\RRR_w$ are calculated from \eqref{eq:wallforce} and \eqref{eq:walltorque} to be
\al{
\RRR_{w,11} &= \RRR_{w,22} = \frac{2}{h} \int_S dS; \label{eq:wallcomp1}\\
\RRR_{w,12} &= \RRR_{w,21} = 0 ;\\
\RRR_{w,13} &= \RRR_{w,31} = \frac{2}{h} \int_S dS \ (-y) ;\label{eq:wallcomp3} \\
\RRR_{w,23} &= \RRR_{w,32} = \frac{2}{h}\int_S dS \ x; \label{eq:wallcomp4} \\
\RRR_{w,33} &= \frac{2}{h} \int_S dS \ (x^2+y^2). \label{eq:wallcomp5}
}
Here, the off-diagonal components of the symmetric tensor $\RRR_w$ are 
explicit manifestations of hydrodynamic rotation-translation coupling for 
anisotropic particles, which vanish for particles with enough symmetry 
\cite{uspal2013engineering}.

Taking Equations \eqref{eq:fluideom} and \eqref{eq:walleom} together, we find 
the same force balance that was obtained above, but with an explicit 
specification of the fluid and wall contributions:
\al{
\left(\begin{array}{c}\bm{F} \\ T\end{array}\right) = -\eta \RRR \left(\begin{array}{c}\bm{U}_p \\ \omega_p\end{array}\right) + \left(\begin{array}{c}\bm{F}_0 \\ T_0 \end{array}\right) = \left(\begin{array}{c}\bm{0} \\ 0 \end{array}\right), \label{eq:vanishingeom}
}
with $\RRR = \RRR_f + \RRR_w$ the symmetric $3 \times 3$ overall resistance 
tensor. The accuracy of this equality is directly related to 
the accuracy of the assumptions underlying the Brinkman equation and the 
simple shear flow in the gaps. Notice that Eq. \eqref{eq:vanishingeom} is 
equivalent to Eq. \eqref{eq:eomsol3D}, indicating that this result is not 
sensitive to the hydrodynamic model one chooses for the hydrodynamic 
fluid-particle interaction, because of the overdamped nature of the 
system and the fact that there are no external force and torque acting on the 
particle.

\subsection{Numerical methods} \label{sec:numericalmethods}
The Stokes equation \eqref{eq:stokes} or the Brinkman equation 
\eqref{eq:brinkman} can be solved numerically for a particle of any shape in 
the channel. In this work, we use the finite-elements software COMSOL 
Multiphysics (COMSOL, Inc., Burlington, MA, USA) to find the flow 
field $\bm{u}$ or $\bar{\bm{u}}$. Using this solution, we can determine the 
forces acting on the particle and integrate its motion using a numerical scheme 
detailed below.

At a given particle position $(x_p,y_p)$ and orientation $\theta$, each column of the resistance tensor $\RRR$ is determined by imposing a single non-zero component of the particle velocities $(\bm{U}_p,\omega_p)$, without external flow. The force and torque that determine the column of $\RRR$ are either found by numerically solving the Stokes equation \eqref{eq:stokes} and integrating the obtained stress tensor via Eq. \eqref{eq:hydroforce}, or by solving the Brinkman equation \eqref{eq:brinkman} and integrating via Eq. \eqref{eq:brinkmantorque} and Eqs. \eqref{eq:wallcomp1} - \eqref{eq:wallcomp5}. Similarly, $\bm{F}_0$ and $T_0$ are found by fixing the particle in the external flow, in either formalism.

With $\RRR, \bm{F}_0$ and $T_0$ determined from the finite elements solutions, the force- and torque-free velocity $\bm{U}_p(x,y,\theta)$ and angular velocity $ \omega_p(x,y,\theta)$ are calculated from Eq. \eqref{eq:vanishingeom}. Then, the particle position and orientation are integrated as
\al{
x(t + \Delta t) &= x(t) +  U_{p,x}(x(t),y(t),\theta(t)) \, \Delta t \label{eq:xdiffeq} \\
y(t + \Delta t) &= y(t) +  U_{p,y}(x(t),y(t),\theta(t)) \, \Delta t  \label{eq:ydiffeq}\\
\theta(t + \Delta t) &= \theta(t) +  \omega_{p}(x(t),y(t),\theta(t)) \, \Delta t, \label{eq:thdiffeq}
}
for some appropriately chosen time step $\Delta t$. The inversion of $\RRR$ to obtain the solution of Eq. \eqref{eq:vanishingeom}, and the numerical integration 
\eqref{eq:xdiffeq} - \eqref{eq:thdiffeq} are performed in MATLAB to obtain the 
position and orientation at the next time step, which are subsequently fed 
back into the finite-elements calculations.

\subsection{Validity of the Brinkman equation}
\begin{figure}[tbp]
   \centering
   \includegraphics[width=0.45\textwidth]{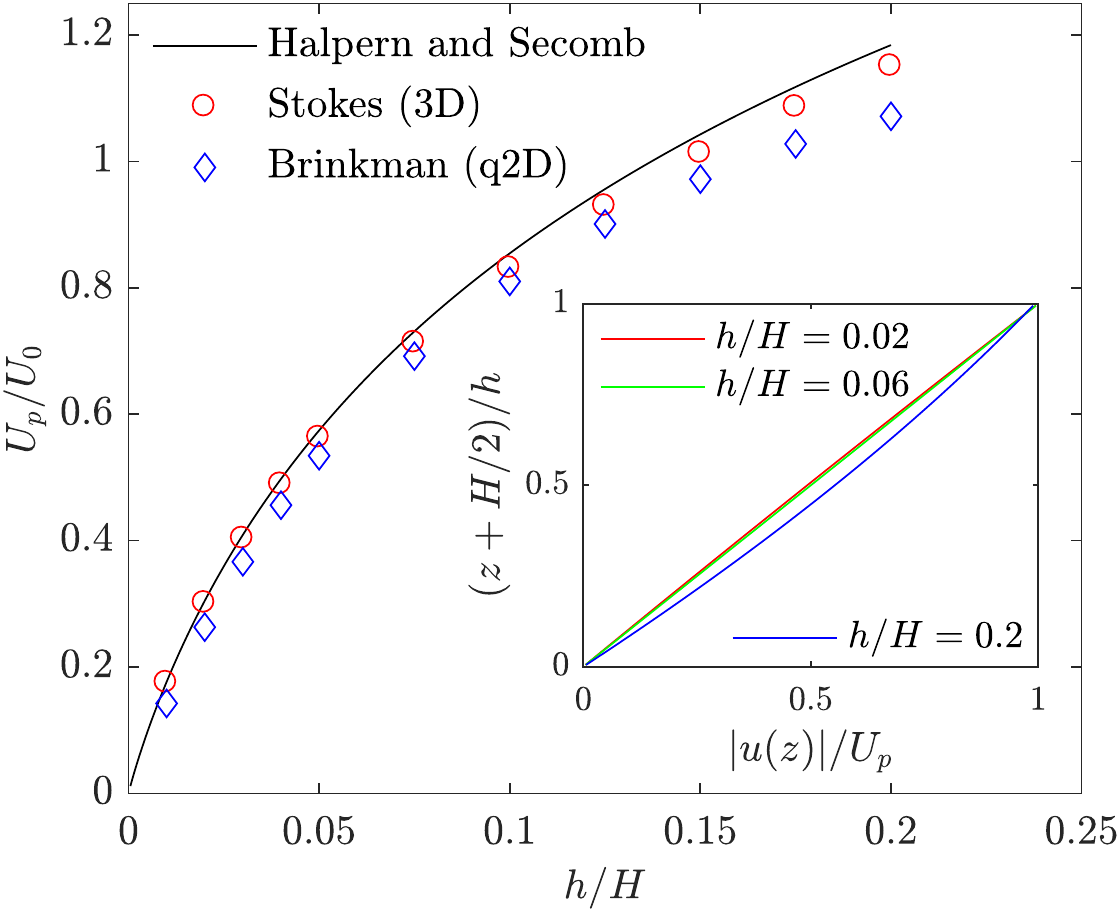} 
   \caption{Terminal velocity of a rounded cylinder between two confining plates. The radius of the cylinder is $R/H =  1.79(1-2 h/H) + r_c$, with radius of curvature $r_c = (H-2h)/2$ for the rounded edges, where $H$ and $h$ are the channel and gap height, respectively. The inset shows the $z$-dependence of the flow field magnitude $|u(z)|$ relative to the particle velocity $U_p$, at position $x=y=0$ in the thin gap between the bottom wall (at $z = -H/2$) and the particle disk surface (at $z = -H/2 + h$), for different gap heights $h$.}
   \label{fig:HS}
\end{figure}

To test the validity of the Brinkman formalism described above, we compare with 
results obtained from three-dimensional analytical or numerical solutions of the 
Stokes equation. Here, we do this by considering the terminal velocity of a disk 
that is moving with the fluid between two infinite parallel plates 
(corresponding to the experimental setup but with $W$ large compared to the disk 
size), for which an analytical result by Halpern and Secomb exists 
\cite{halpern1991viscous}. The analysis in Ref. \cite{halpern1991viscous} concerns a cylindrical disk with 
rounded edges with a radius of curvature that is precisely half the height of 
the cylinder. In. Fig. \ref{fig:HS}, we plot the  
particle velocity $U_p$, scaled by $U_0$, as a function of the gap height $h/H$. We 
compare the analytical solution to the results obtained from the solutions of 
the Stokes equation and of the Brinkman equation. We find good agreement 
between the three calculations, especially for smaller gaps. The deviations at 
larger gaps are expected, since the assumption of a simple shear flow and the 
quasi-2D character ceases to hold in this region. To test this, we plot in the 
inset of Fig. \ref{fig:HS} the magnitude of the flow field in the thin gap 
between the bottom wall at $z = -H/2$, where we have $|\bm{u}| = 0$, and the 
disk surface at $z = -H/2 + h$, where $|\bm{u}| = U_p$. We find that, for $h/H 
= 0.02$, the dependence of the flow field on the height $z$ is linear, 
corresponding to simple Couette flow. Conversely, for $h/H = 0.2$, we find a 
deviation from the linear dependence, indicating that the simple shear flow 
assumption ceases to hold. The case of $h/H = 0.06$ corresponds to the 
experiments in this work, where from Fig. 
\ref{fig:HS} we see that the simple shear flow assumption is still valid.

It should be noticed that the rounding of cylinder edges is not taken into 
account in the quasi-2D calculations. This does not seem to influence 
the results strongly, since the results agree with the analytical and 
three-dimensional numerical results where the rounded edges are taken into 
account. An attempt to incorporate rounded edges, by making the gap height 
in our quasi-2D calculations position-dependent, did not improve the results 
significantly, but increased the computational time by a factor two and was 
therefore not continued.

\section{Results}
To test our numerical scheme for complex particle geometries, where analytical solutions are not available, we turn our attention to the situation described by Uspal {\it et al.} \cite{uspal2013engineering}. In Ref. \cite{uspal2013engineering} and in this work, dumbbell-shaped particles are produced in the channel using `continuous flow lithography'. These dumbbell particles consist of two circular disks of radius $R_1$ and $R_2\le R_1$, respectively, and height $H - 2h$. The smaller disk has a fixed radius $R_2 = \SI{18.75}{\micro \meter}$, while the larger disk radius $R_1$ is varied. The disks are connected, at a fixed center-to-center distance $s = \SI{62.5}{\micro \meter}$, by a cuboid of width $\SI{13.7}{\micro \meter}$, which has the same height as the two disks. In the experiments, the dumbbell edges are not completely sharp but are rounded, with a rounding radius $r_c/H = 0.15$, as estimated from experimental images \cite{uspal2013engineering}. This detail is incorporated in the three-dimensional calculations, but not in the two-dimensional calculations. The particle geometry is depicted in Fig. \ref{fig:geom}b.

In this work, we focus on the aligning motion observed in Ref. \cite{uspal2013engineering}, where asymmetric dumbbells oriented in the flow with the larger disk upstream. The orientation $\theta$ between the long axis of the dumbbell and the flow direction (see Fig. \ref{fig:geom}a) was shown to follow the equation
\eqn{ \label{eq:taueq}
\dot{\theta} = - \frac{1}{\tau} \sin \theta,
}
where the characteristic timescale $\tau$ is dependent on the particle geometry. Specifically, we investigate here the dependence of $\tau$ on the ratio of the disk radii $R_1/R_2$. From Eq. \eqref{eq:taueq}, we see that $\tau$ can be obtained from a trajectory by considering the angular velocity at perpendicular orientation:
\eqn{ \label{eq:taueq2}
\tau = - \frac{1}{\dot{\theta} ( \theta = \pi/2)}.
} 
Hence, we can use the 3D and quasi-2D calculations described above to find the angular velocity of a particular dumbbell and obtain $\tau$ through \eqref{eq:taueq2}. 

\begin{figure}[tbp]
   \centering
   \includegraphics[width=0.45\textwidth]{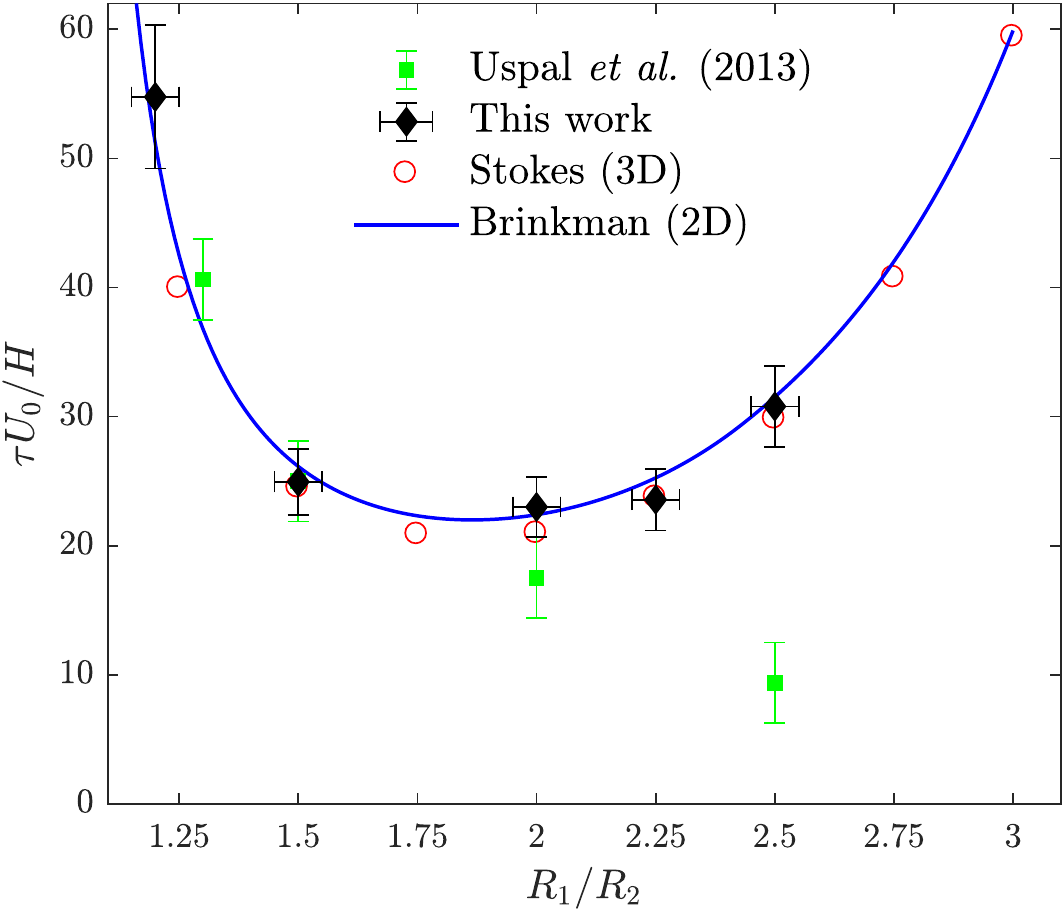} 
   \caption{The characteristic reorientation time $\tau$ for dumbbell particles, as obtained from 3D finite-elements method in the Stokes formalism (red) and 2D finite-elements in the Brinkman formalism (blue), and experimental data from the previous experiments of Ref. \cite{uspal2013engineering} (green) and new experiments (black), as a function of $R_1/R_2$.}
   \label{fig:tau}
\end{figure}

The results are shown in Fig. \ref{fig:tau}, where we plot $\tau$ against $R_1/R_2$. In green squares, we show the experimental data of \cite{uspal2013engineering}; the red circles show the numerical data from our three-dimensional calculations, while the blue line shows the results from the quasi-two-dimensional calculations. Firstly, we observe excellent agreement between the numerical results, which serves as another confirmation of the validity of our quasi-two-dimensional calculations. Secondly, we see that the results agree with the previous experimental data, with the exception of the data point of the most asymmetric size ratio $R_1/R_2 = 2.5$. Our calculations clearly show that a minimum in $\tau$ is expected around $R_1/R_2 \approx 1.9$, with $\tau$ increasing for larger $R_1/R_2$. 

In contrast, the previous data show a strictly decreasing dependence of $\tau$ on $R_1/R_2$. Moreover, the theoretical curve that was fitted to the data in Ref. \cite{uspal2013engineering} was monotonically decreasing. This curve was derived considering an idealized situation of a pair of disks connected with a massless rod, which demonstrates the pitfalls of oversimplifying the geometry . However, we argue here that the minimum for $\tau$ that is predicted by our analysis is correct. In our calculations, as well as in the experiments, $R_1/R_2$ is varied by varying $R_1$, while keeping $R_2$ and the center-to-center distance $s$ constant. As a result, when $R_1 \geq R_2 + s$, the shape becomes effectively just a single disk, which will not rotate at all ($\tau \to \infty$) due to symmetry (provided it is still in the center of the channel). Therefore, when $R_1/R_2 \to \infty$, we should have $\tau \to \infty$ and hence a minimum in $\tau$. Note that the overlapping of the two disks is not incorporated in the theoretical results of Ref. \cite{uspal2013engineering}.

To clear up this discrepancy, we performed a new set of experiments using an experimental setup that is almost identical to the setup in Ref. \cite{uspal2013engineering}. Dumbbell particles were produced in a Hele-Shaw channel using continuous flow lithography and their reorientation motion was tracked, as described in Appendix \ref{app:exp}. In our experimental setup, the particle geometry and gap heights are taken to be approximately identical to those used in Ref. \cite{uspal2013engineering}. Our measured reorientation times $\tau$ are shown by black diamonds in Fig. \ref{fig:tau}, where an excellent agreement with our numerical results is observed. The new data confirms the existence of a minimum in the $\tau$ as a function of $R_1/R_2$. We stress that our numerical method does not rely on any adjustable parameter, and uses only the experimental geometry and flow rate as input.

\begin{figure}[tbp]
   \centering
   \includegraphics[width=0.5\textwidth]{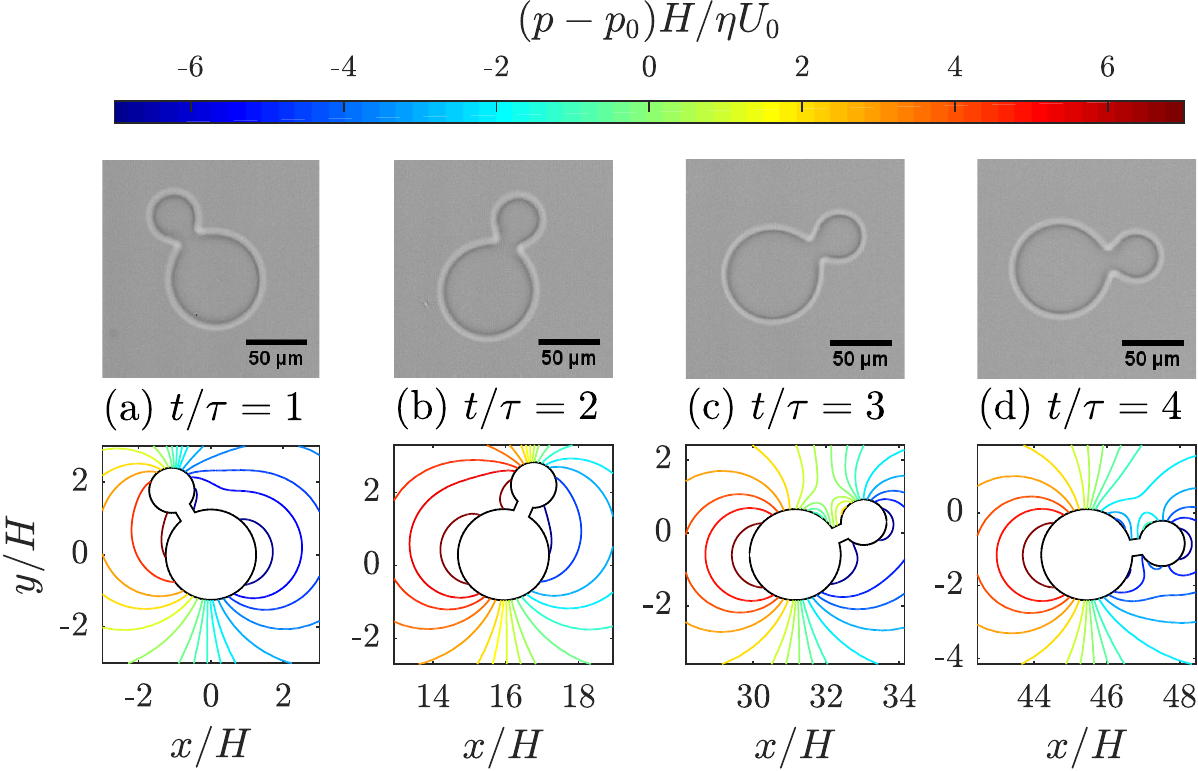} 
   \caption{Experimental (top) and numerical (bottom) snapshots of the reorienting motion of a dumbbell with $R_1/R_2 =2$ in the microfluidic channel, at $t/\tau = 1$ (a), $t/\tau = 2$ (b), $t/\tau = 3$ (c), $t/\tau = 4$ (d), where the initial orientation of the dumbbell is $\theta(0) = 5\pi/6$. We clearly observe that the dumbbell reorients with the larger disk upstream ($\theta =0$). Around the dumbbell we show isobars of the disturbance pressure field created by the particle, as obtained from Eq. \eqref{eq:brinkman}.} \label{fig:stills}
\end{figure}

We have also calculated the complete angular trajectory of the dumbbell particles. Setting the time step to $\Delta t /\tau = 0.2$, we integrated the position and orientation as described in Eqs. \eqref{eq:xdiffeq} - \eqref{eq:thdiffeq}, starting from an initial orientation $\theta(t=0) = 5\pi/6$. Indeed, we observe that the dumbbells will align with the flow such that the larger disk is upstream ($\theta = 0$). This is illustrated in Fig. \ref{fig:stills}, where we contrast experimental (top) and numerical (bottom) snapshots of the reorienting particle at times $t/\tau = 1$ (a), $t/\tau = 2$ (b), $t/\tau = 3$ (c), and $t/\tau = 4$ (d). We do not distinguish here between the $\tau$ values obtained from experiment and numerical calculations, since the two results are found to agree very well. Around the particle in the numerical snapshots of Fig. \ref{fig:stills}, we show isobars of the disturbance pressure field $p(x,y)-p_0(x)$ created by the particle, which are obtained from the Brinkman equation \eqref{eq:brinkman} with the force- and torque-free (angular) velocity imposed. Here, $p_0$ denotes the pressure field corresponding to an undisturbed external flow $U_0$ in the channel. The contours of $p(x,y)-p_0(x)$ show the dipolar nature of the disturbance flow. A clear impression of the particle motion may be obtained from the microscopic movie and animations found in the supplemental material. \footnote{See the ancillary files of this preprint for the animations and microscopic movie of a reorienting dumbbell particle. Descriptions of these animations are found below in the Supplemental material.}

\begin{figure}[tbp]
   \centering
   \includegraphics[width=0.45\textwidth]{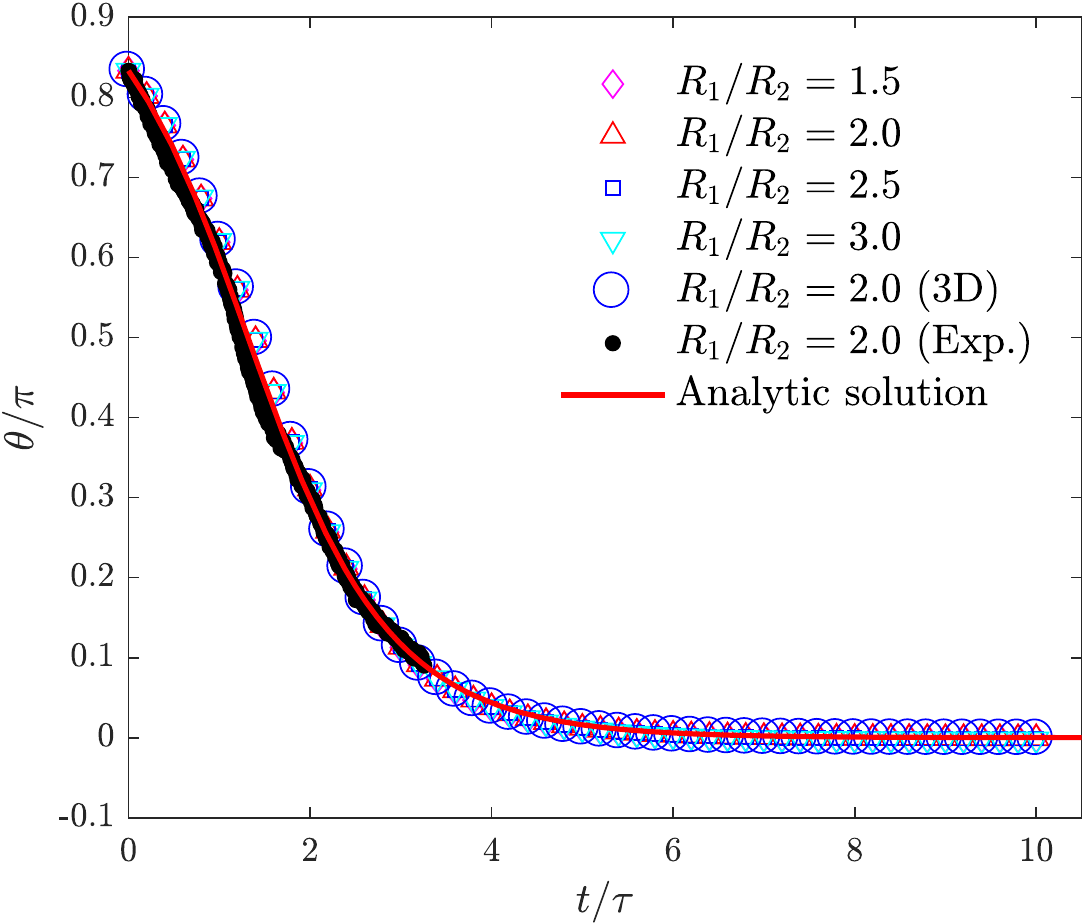} 
   \caption{The orientation angle $\theta$ as a function of rescaled time $t/\tau$, with $\theta(t=0) = 5 \pi /6$, for dumbbells of varying $R_1/R_2$. The large blue circles show the trajectory obtained using the three-dimensional method, while the black dots show one particular trajectory obtained from the experiments, both for a dumbbell particle with $R_1/R_2 = 2$. Errorbars for the experimental data are smaller than the black dots and hence omitted. The solid red line shows the analytical solution of equation \eqref{eq:taueq}.}
   \label{fig:theta_t}
\end{figure}

The results for the complete angular trajectories are shown in Fig. \ref{fig:theta_t}. When time is rescaled by $\tau$, we find a data collapse of the numerical calculations (open symbols) on the analytical solution of Eq. \eqref{eq:taueq} (solid red line). This collapse offers a strong confirmation that the orientation is indeed correctly described by equation Eq. \eqref{eq:taueq}. For $R_1/R_2 = 2.0$, we have also calculated a complete trajectory using the three-dimensional method. The results, shown by the large blue circles in Fig. \ref{fig:theta_t}, perfectly agrees with the quasi-two-dimensional results, a final confirmation of the accuracy of our quasi-two-dimensional calculations. Finally, we also show an experimentally obtained trajectory (black dots), and find good agreement with the numerical results. 
Discrepancies may be attributed to small variations in the channel height due to fabrication imperfections or dust particles in the pre-polymer mixture.

\section{Summary and Outlook}
In conclusion, we have set up a combined theoretical and numerical framework that uses numerical solutions of either three-dimensional (Stokes) or quasi-two-dimensional (Brinkman) hydrodynamical equations, to calculate strongly confined particle motion in shallow microfluidic channels. Our method is not restricted to simplified shapes such as disks, but is able to handle any shape. The method is validated by comparing between analytical and three- and quasi-two-dimensional numerical calculations, which shows excellent agreement. The two orders of magnitude of computational speed-up that is offered by the quasi-two-dimensional description, enables to fully resolve the particle trajectory in time.

Our method is applied to dumbbell particles, for which we have calculated the characteristic rotation time $\tau$ as a function of $R_1/R_2$, and found that, contrary to earlier findings, $\tau$ shows a minimum at $R_1/R_2 \simeq 1.9$. The existence of a minimum is confirmed by a new set of experiments. Moreover, we have calculated the angular motion as a function of time, and shown that it agrees with the equation of motion derived earlier, and with the trajectories that are obtained from the experiments.

In future work, it will be interesting to further investigate the relation between the geometry and the trajectories of different particles, with the goal to possibly steer particles to different areas in the channel.  Lastly, our method is easily generalized to systems of multiple particles, which can offer a controlled setup to study hydrodynamic interaction between confined particles. Work in this direction is already under-way.  

\begin{acknowledgments}
This work is part of the D-ITP consortium, a program of the Netherlands Organisation for Scientific Research (NWO) that is funded by the Dutch Ministry of Education, Culture and Science (OCW). We acknowledge financial support from an NWO-VICI grant. S.S acknowledges funding from the 
European Union's Horizon 2020 programme under the Marie 
Sk\l{}odowska-Curie grant agreement No. 656327. We thank A. Wijkamp of the van 't Hoff Laboratory for Physical and Colloid Chemistry for providing us with fluorescently-labelled polystyrene beads and S. O. Toscano for her help in acquiring experimental results.
\end{acknowledgments}

\appendix

\section{Materials \& Experimental Methods} \label{app:exp}
\subsection{Particle production \& tracking}
Polymeric microparticles are produced and observed with an experimental setup, similar to the one used by Uspal \textit{et al.} \cite{uspal2013engineering}. Polydimethylsiloxane (PDMS, Sylgard$^\textrm{\textregistered}$ 184, Dow Corning), microfluidic devices are produced according to \cite{dendukuri2008pdms}. A UV-crosslinking oligomer, poly-(ethyleneglycol) diacrylate (PEG-DA $M_n=700$, $\eta = 
\SI{55e-3}{\pascal \second}$, Sigma-Aldrich), is mixed with a photoinitiator, hydroxy-2-methylpropiophenone, (Darocur$^\textrm{\textregistered}$ 1173, Sigma-Aldrich), in a 19:1 volume ratio and the mixture is pumped through the microfluidic channel. The device, loaded with prepolymer, is mounted on the stage of a motorized Nikon Ti Eclipse inverted optical microscope. A photolithographic mask with well-defined shape is inserted between the UV light source and the microscope objective. Mask designs are made in Wolfram Mathematica$^\textrm{\textregistered}$ and post-processed in Dassault Syst{\'e}mes' DraftSight$^\textrm{\textregistered}$.

Microparticles are produced by shining a 100 ms pulse of UV light through the mask onto the channel, thus confining photopolymerization to a discrete part of the prepolymer mixture. Oxygen, diffusing through the PDMS walls of the device, inhibits crosslinking \cite{dendukuri2008pdms}. This ensures the formation of two lubrication layers, which separate the particle from the \textit{z}-walls, thus preventing sticking. 

The microparticle is set in motion by applying a pressure drop $\Delta p$ across the channel and tracked by moving the automated stage in a stepwise manner. Since two different microfluidic devices with  differing hydraulic resistences were used, the applied pressure  differs depending on the device used \--- $\Delta p_1=\SI[separate-uncertainty=true,multi-part-units=single]{0.10(1)}{psig}$ and $\Delta p_2=\SI[separate-uncertainty=true,multi-part-units=single]{0.15(1)}{psig}$. Both pressure drops produce a flow with $U_0\approx\SI{50}{\micro\metre\per\second}$ in the respective microfluidic device. The experimental uncertainty in determining $U_0$ is around $10\%$ \footnote{See Supplemental Material below for detailed calculation of all reported experimental uncertainties.}. 

Particle positions and orientations are extracted from the acquired movie using a custom-written MATLAB script, which employs circular Hough transforms to identify the particle shape in each frame. The script utilizes MATLAB's Bio-Formats package \cite{Linkert2010} and the calcCircle tool.
\subsection{Channel geometry \& determination}
Both used devices feature a microfluidic channel of length ($L=\SI[separate-uncertainty=true,multi-part-units=single]{11.54(1)}{\centi\meter}$) and width ($W=\SI[separate-uncertainty=true,multi-part-units=single]{515(2)}{\micro\meter}$) determined via optical microscopy. The channel height ($H=\SI[separate-uncertainty=true,multi-part-units=single]{30(1)}{\micro\meter}$) is inferred from fluorescent particle tracking \cite{uspal2013engineering} and verified with optical and scanning electron microscopy by cutting the device and looking at the cross section.
Briefly, fluorescently-labelled polystyrene beads with an average diameter of $\SI{1.89}{\micro\metre}$ are introduced in the prepolymer mixture. After focussing in the mid-plane of the channel, the beads are tracked, as the mixture is pumped through the device. Due to optical depth of focus and bead size, a normal distribution of particle velocities is obtained. The mean of the distribution is taken as $U_0$ and the height of the channel is calculated via Eq. \ref{eq:height}.  
Particle height ($H_\text{p}=\SI[separate-uncertainty=true,multi-part-units=single]{25.7(14)}{\micro\meter}$) is measured via optical microscopy and is used to calculate the lubrication layer thickness ($h=(H-H_\text{p})/2=\SI[separate-uncertainty=true,multi-part-units=single]{2.2(9)}{\micro\meter}$). 

\section{Lorentz reciprocal theorem for Brinkman flow} \label{app:lorentz}
In this appendix, we derive a version of the Lorentz reciprocal theorem for solutions of the Brinkman equation. Let $\bar{\bm{u}}$ and $\bar{\bm{u}}'$ be two solutions to the Brinkman equation in a two-dimensional fluid domain $A$ with boundary $\del A$, with different boundary conditions. These solutions have corresponding stress tensors $\bar{\sigma}$ and $\bar{\sigma}'$, defined as
\eqn{
\bar{\sigma}_{ij} = - \bar{p} \delta_{ij} + \eta H (\del_i \bar{u}_j + \del_j \bar{u}_i),
}
and similarly for $\bar{\sigma}'$. With this definition, we may write the Brinkman equation \eqref{eq:brinkman} as
\eqn{ \label{eq_ihs:brinkmanrewritten}
\nabla_{\text{2D}} \cdot \bar{\sigma} = \frac{12 \eta}{H} \bar{\bm{u}}, \quad \nabla_{\text{2D}} \cdot \bar{\bm{u}} = 0.
}
Let us follow the same steps as in the derivation of the original Lorentz reciprocal theorem \cite{kim1991}. First, consider the quantity $\bar{u}'_i(\del_j \bar{\sigma}_{ij})$, which can be rewritten as
\al{
\bar{u}'_i(\del_j \bar{\sigma}_{ij}) &= \del_j(\bar{u}'_i \bar{\sigma}_{ij}) - 
\bar{\sigma}_{ij} \del_j \bar{u}'_i = \del_j(\bar{u}'_i \bar{\sigma}_{ij}) - (-p 
\delta_{ij} + \eta (\del_{i} \bar{u}_{j}+ \del_j \bar{u}_i))\del_j\bar{u}'_i 
\nonumber \\
&=\del_j(\bar{u}'_i \bar{\sigma}_{ij}) -  \eta (\del_{i} \bar{u}_{j}+ \del_j \bar{u}_i)\del_j\bar{u}'_i,
}
where the term involving $\bar{p}$ drops out due to the incompressibility constraint $\del_i \bar{u}'_i = 0$. Interchanging $\bar{\bm{u}}$ and $\bar{\bm{u}}'$ and subtracting gives us
\al{
\label{eq:lorenz_step1}
\bar{u}'_i (\del_j \bar{\sigma}_{ij})  - \bar{u}_i(\del_j \bar{\sigma}'_{ij})   
= \del_j(\bar{u}'_i \bar{\sigma}_{ij} - \bar{u}_i \bar{\sigma}'_{ij}). 
}
For solutions of Stokes flow, both terms of the left-hand-side of Eq.
\eqref{eq:lorenz_step1} vanish identically leading to the Lorentz reciprocal 
theorem. For solutions of the Brinkman equation, the 
left-hand-side of Eq. \eqref{eq:lorenz_step1} vanishes because when Eq.
\eqref{eq_ihs:brinkmanrewritten} is applied to it, we find
\eqn{
\bar{u}'_i \frac{12 \eta}{H} \bar{u}_i - \bar{u}_i \frac{12 \eta}{H} \bar{u}'_i = 0,
}
such that Eq. \eqref{eq:lorenz_step1} reads
\eqn{
\nabla_{\text{2D}} \cdot (\bar{\bm{u}}' \cdot \bar{\sigma} - \bar{\bm{u}} \cdot \bar{\sigma}') = 0.
}
Thus, using the two-dimensional divergence theorem, we obtain the Lorentz reciprocal theorem for Brinkman flow:
\eqn{
\oint_{\del A} ds \bar{u} \cdot \bar{\sigma}' \cdot \bm{n} = \oint_{\del A} ds \bar{u}' \cdot \bar{\sigma} \cdot \bm{n},
}
where $\del A$ denotes the one-dimensional boundary of the two-dimensional fluid domain $A$ where $\bar{\bm{u}}$ and $\bar{\bm{u}}'$ are defined and solve the Brinkman equation.Using this relation, symmetry properties of the resistance tensor $\RRR_f$ 
defined in \eqref{eq:fluideom} for particles in Hele-Shaw cells may be proven, 
see Appendix \ref{app:brinkmanresistance}.

\section{Fluid resistance tensor from the Brinkman equation} \label{app:brinkmanresistance}
In this appendix, we show explicitly that the force $\bm{F}_f$ and torque $T_f$ exterted on the particle by the (quasi-)two-dimensional Brinkman fluid are related to the particle velocity and angular velocity in terms of a $3 \times 3$ resistance tensor $\RRR_f$.
Inspired by the decompositions into sub-solutions of the original Stokes problem in Section \ref{sec:stokeseom}, we exploit the linearity of the Brinkman equation to write the solutions as the superpositions $\bar{\bm{u}} = \bar{\bm{u}}_0 + \bar{\bm{u}}' + \bar{\bm{u}}''$ and $\bar{p} = \bar{p}_0 + \bar{p}' + \bar{p}'' $, where $(\bar{p}_0,\bar{\bm{u}}_0)$, $(\bar{p}',\bar{\bm{u}}')$ and $(\bar{p}'',\bar{\bm{u}}'')$ are solutions to the Brinkman equation \eqref{eq:brinkman} with boundary conditions
\al{
&\bar{\bm{u}}_0\big|_{\text{walls}} = 0,  &&\bar{\bm{u}}_0\big|_{\text{inlet}} = U_0 \hat{\bm{x}},   &&\bar{\bm{u}}_0\big|_{\bm{r} \in \del \bar{S}_p} = 0;  \\
&\bar{\bm{u}}'\big|_{\text{walls}} = 0, &&\bar{\bm{u}}'\big|_{\text{inlet}} = 0,  &&\bar{\bm{u}}'\big|_{\bm{r} \in \del \bar{S}_p} = \bm{U}_{p}; &\label{eq:brinkmanbctranslating}\\
&\bar{\bm{u}}''\big|_{\text{walls}} = 0, &&\bar{\bm{u}}''\big|_{\text{inlet}} = 0,  &&\bar{\bm{u}}''\big|_{\bm{r} \in \del \bar{S}_p} =  \omega_p(-(y-y_p),x-x_p),\label{eq:brinkmanbctrotating}
}
respectively. Here, $\bar{\bm{u}}_0$ corresponds to the situation in which the particle is stationary subject to the imposed external flow, while $\bar{\bm{u}}'$ and $\bar{\bm{u}}''$ correspond to the situation where the particle is translating and rotating, respectively, in the absence of an imposed flow. 
The stress tensor of Eq. \eqref{eq:brinkmanstress}, being linear in the fields $\bar{\bm{u}}$ and $\bar{p}$, admits a similar decomposition $\bar{\sigma} = \bar{\sigma}_0 + \bar{\sigma}' + \bar{\sigma}''$, which is then inherited by the force $\bm{F}_f$ and torque $T_f$ via Eq. \eqref{eq:brinkmantorque}. While the force and torque on the stationary particle follow directly from Eq. \eqref{eq:brinkmantorque} by substituting $\bar{\sigma}_0$ for $\bar{\sigma}$, let us consider the contributions from $\bar{\sigma}'$ and $\bar{\sigma}''$ in more detail.

\subsubsection{Translation}
The linearity of the Brinkman equation allows for a factorization 
\eqn{ \label{eq:factorbrinkmantrans}
\bar{u}'_i = \bar{\mathcal{U}}'_{ij} U_{p,j}, \quad \bar{p}' = \eta \bar{\mathcal{P}}'_{ij} U_{p,j},
}
where the tensor fields $(\bar{\mathcal{U}}',\bar{\mathcal{P}}')$ obey
\al{
&- \del_i \bar{\mathcal{P}}'_j + \del_k^2 \bar{\mathcal{U}}'_{ij} -\frac{12 }{H} \bar{\mathcal{U}}'_{ij}  = 0, \quad \del_i \mathcal{U}'_{ij} = 0, \\
&\bar{\mathcal{U}}'_{ij}\big|_{\text{walls}} = \bar{\mathcal{U}}'_{ij}\big|_{\text{inlet}} = 0, \quad \bar{\mathcal{U}}_{ij}'\big|_{\bm{r} \in \del \bar{S}_p} = \delta_{ij},
}
which lead to the original Brinkman equation for $(\bar{\bm{u}}',\bar{p}')$ by contracting with $\bm{U}_p$. In turn, the fields $(\bar{\mathcal{U}}',\bar{\mathcal{P}}')$ define a tensor
\eqn{ \label{eq:brinkmantransgenstresstensor}
\bar{\Sigma}'_{ijk} = - \delta_{ij} \bar{\mathcal{P}}'_k + \del_i \bar{\mathcal{U}}'_{jk} + \del_j \bar{\mathcal{U}}'_{ik},
}
which is related to the stress tensor by $\bar{\sigma}'_{ij} = \eta \bar{\Sigma}'_{ijk} (U_{p})_k$. Next, we define 
\al{
\bar{K}_{ik} &= - \oint_{\del \bar{S}_p} dS (\bar{\Sigma}'_{ijk} n_j), \label{eq:Kdef} \\
\bar{C}'_{k} &= -\oint_{\del \bar{S}_p} dS \bigg( (x-x_p) (\bar{\Sigma}'_{y j k} n_j) -  (y-y_p) (\bar{\Sigma}'_{x j k} n_j)   \bigg), \label{eq:Cprimedef}
}
such that the hydrodynamic force and torque that the two-dimensional fluid exerts on the translating particle are expressed as 
\eqn{
F'_{f,i} = - \eta \bar{K}_{ij} {U}_{p,j} \ , \quad T'_f = - \eta \bar{C}'_j {U}_{p,j} \ . \label{eq:kforce} 
}

Similar to the the derivation for the three-dimensional resistance tensor, we can prove that the $2 \times 2$ tensor $K$ is symmetric, using a version of the Lorentz reciprocal theorem that is proven in Appendix \ref{app:lorentz}. Consider the two problems where the particle is either translating with velocity $\bm{U}$ or $\tilde{\bm{U}}$ through the channel without external flow, with corresponding flow fields $\bar{\bm{u}}$ and $\tilde{\bar{\bm{u}}}$ that are solutions to the Brinkman equation with boundary conditions \eqref{eq:brinkmanbctranslating}. Applying the Brinkman version of the reciprocal theorem, we find
\al{
\oint_{\del \bar{S}_p} ds \bar{\bm{u}} \cdot \tilde{\bar{\sigma}} \cdot \bm{n} + \oint_{\text{w,i,o}} ds \bar{\bm{u}} \cdot \tilde{\bar{\sigma}} \cdot \bm{n} = \oint_{\del \bar{S}_p} ds \tilde{\bar{\bm{u}}} \cdot {\bar{\sigma}} \cdot \bm{n} + \oint_{\text{w,i,o}} ds \tilde{\bar{\bm{u}}} \cdot {\bar{\sigma}} \cdot \bm{n},
}
where the integration over `$\text{w,i,o}$' is carried out over the (projected) sidewalls, the inlet and the outlet. However, since both flow fields vanish on the side walls according to Eq. \eqref{eq:brinkmanbctranslating}, these integrals vanish identically. Using the no-slip boundary condition on the particle boundary $\del \bar{S}_p$ and definitions of Eq. \eqref{eq:brinkmantorque}, we find
\eqn{
\tilde{\bm{U}} \cdot \bm{F}_f = {\bm{U}} \cdot \tilde{\bm{F}}_f,
}
which using Eq. \eqref{eq:kforce} may be written as
\eqn{
\tilde{U}_i \bar{K}_{ij} U_j = U_i \bar{K}_{ij} \tilde{U}_j =  \tilde{U}_i \bar{K}_{ji}  U_j. 
}
Now, since $\bm{U}$ and $\tilde{\bm{U}}$ are completely arbitrary, we conclude that $\bar{K}$ is symmetric: $\bar{K}_{ij} = \bar{K}_{ji}$.

\subsubsection{Rotation}
Similarly, we factorize the angular velocity $\omega_p$ from the solution $\bar{\bm{u}}''$:
\eqn{ \label{eq:factorbrinkmanrot}
\bar{u}''_i = \bar{\mathcal{U}}''_{i} \omega_p, \quad \bar{p} = \eta \bar{\mathcal{P}}'' \omega_p,
}
with (tensor-)fields that obey 
\al{
&- \del_i \bar{\mathcal{P}}'' + \del_k^2 \bar{\mathcal{U}}''_{i} -\frac{12 }{H} \bar{\mathcal{U}}''_{i}  = 0, \quad \del_i \mathcal{U}''_{i} = 0, \\
&\bar{\mathcal{U}}''_{i}\big|_{\text{walls}} = \bar{\mathcal{U}}''_{i}\big|_{\text{inlet}} = 0, \quad \bar{\mathcal{U}}''_i\big|_{(x,y) \in \del \bar{S}_p} = -y \delta_{ix } + x \delta_{i y},
}
which, as usual, lead to the original Brinkman equation for $\bar{\bm{u}}''$ with boundary conditions \eqref{eq:brinkmanbctrotating} by multiplying with $\omega_p$. Moreover, we define the tensor $\bar{\Sigma}''_{ij}$, similar to Eq. \eqref{eq:brinkmantransgenstresstensor}, which has the property $\bar{\sigma}_{ij}'' = \bar{\Sigma}''_{ij} \omega_p$. Note that the fields $\bar{\mathcal{U}}'',\bar{\mathcal{P}}''$ and $\bar{\Sigma}''$ have dimensions of length of one power higher than their single primed counterparts $\mathcal{U}',\mathcal{P}'$ and $\Sigma'$, while they are of one tensor rank lower, the latter following from the fact that we have only one rotational degree of freedom in two dimensions. To proceed, we define
\al{
\bar{C}''_{i} &= - \oint_{\del \bar{S}_p} dS (\bar{\Sigma}''_{ij} n_j), \\
\bar{\Omega} &= -\oint_{\del \bar{S}_p} dS (x-x_p)(\bar{\Sigma}''_{y j} n_j) - (y-y_p) (\bar{\Sigma}''_{x j} n_j), \label{eq:omegadefbrinkman}
}
such that
\eqn{
\bm{F}''_{f,i} = -\eta \bar{C}''_i \omega_p , \quad T''_f = -\eta \bar{\Omega}\omega_p. \label{eq:omegatorque}
}
Similar to proving that $\bar{K}$ is symmetric, we can use the Lorentz reciprocal theorem of Appendix \ref{app:lorentz} to prove that in fact $C'_i = C''_i$, but we leave this to the reader.

\subsubsection{Complete particle motion}
Taking together the contributions from the sub-solutions $\bar{\bm{u}}_0, \bar{\bm{u}}'$ and $\bar{\bm{u}}''$, we find that the force and torque on the moving particle exerted by the two-dimensional fluid can be written as
\eqn{ \label{eq:fluideomapp}
\left(\begin{array}{c}\bm{F} \\ T\end{array}\right)_f = -\eta \RRR_f \left(\begin{array}{c}\bm{U}_p \\ \omega_p \end{array}\right)  +  \left(\begin{array}{c}\bm{F}_0 \\ T_0 \end{array}\right),
}
where the components of the $3\times3$ resistance tensor $\RRR_f$ are given by
\eqn{
\RRR_f = \left(\begin{array}{ccc}\bar{K}_{11} & \bar{K}_{12} & \bar{C}_1 \\ \bar{K}_{12} & \bar{K}_{22} & \bar{C}_2 \\ \bar{C}_1 & \bar{C}_2 & \bar{\Omega}\end{array}\right),
}
with $C_i \equiv C_i' = C_i''$.
%
%
%

\bibliography{../Bibliography}

\widetext
\clearpage
\begin{center}
\textbf{\large Supplemental material: Calculating the motion of highly confined, arbitrary-shaped particles in Hele-Shaw channels}
\end{center}
\setcounter{equation}{0}
\setcounter{figure}{0}
\setcounter{table}{0}
\setcounter{page}{1}
\setcounter{section}{0}
\makeatletter
\renewcommand{\theequation}{S\arabic{equation}}
\renewcommand{\thesection}{S\arabic{section}}
\renewcommand{\thefigure}{S\arabic{figure}}
\renewcommand{\bibnumfmt}[1]{[S#1]}
\renewcommand{\citenumfont}[1]{S#1}
\section*{Description of the movies}
Here, we give a short description of the experimentally-obtained movie and the animations provided as supplementary information. Definitions of the shape parameters, the channel dimensions and other parameters may be found in the main text.

\begin{enumerate}
%
\item[{\bf M1}] Movie 1 is obtained from microscopic images and shows a dumbbell particle with $R_1/R_2 = 2$ rotating towards a stable orientation with the large disk upstream, while translating in the positive $x$-direction (to the right), along with the externally imposed flow. The field of view of the microscope follows the particle in a stepwise manner.

\item[{\bf A1}] Animation 1 shows a reorienting dumbbell particle with $R_1/R_2 = 2$, as seen from a fixed (laboratory) frame. The color of the heat map indicates the magnitude of the flow field $\bm{u}$ in units of $U_0$.

\item[{\bf A2}]  Animation 2 shows again a reorienting dumbbell particle with $R_1/R_2=2$, now seen from a frame co-moving with the particle. The arrows indicate the disturbance flow field $\bar{\bm{u}} - \bar{\bm{u}}_0$, where $\bar{\bm{u}}_0$ is the externally imposed flow field far away or in the absence of the particle. The colors of the arrows correspond to the magnitude $|\bar{\bm{u}} - \bar{\bm{u}}_0|$ of the disturbance flow field.

\item[{\bf A3}]  Animation 3 shows again a reorienting dumbbell particle with $R_1/R_2=2$, seen from a frame co-moving with the particle, now shown with isobars of the disturbance pressure field $p - p_0$, in units of $\eta U_0 / H$. Here, $p_0$ denotes the externally imposed pressure field far away from the particle.

\end{enumerate}

\section*{Determination of experimental uncertainty}

To calculate the experimental uncertainty in determining the scaled rotation time $\tilde{\tau}=\tau U_0/H$ and the time evolution of the particle orientation with respect to the flow $\theta\pars{t}$, we carry out an error propagation analysis. In the general case of an experimental quantity $F$ dependent on a series of experimentally-determined quantities $\brcs{x_i}$ with uncertainties $\brcs{\sigma_{x_i}}$, one calculates the uncertainty in $F$ as:
\begin{equation}
\sigma_F=\sqrt{\sum_{\brcs{x_i}}\pars{\pder{R\pars{\brcs{x_i}}}{x_i}\sigma_{x_i}}^2}.
\end{equation}

\subsection{Uncertainty in the particle thickness}

The thickness of the particles, $H\ut{p}=H-2h$, is measured at the outlet, where they topple over. The particles are observed from the side and the distance between the edges of the top and bottom face is calculated. Due to uneven illumination of the faces and low contrast of the system, the edges appear thicker than one pixel, therefore a type B uncertainty should be taken into consideration. Type B uncertainty reflects the random nature of a process and any human bias in measuring it. Since most observed edges are roughly 5 pixels thick, a type B uncertainty $\sigma_\text{B}$ of 3 pixels is estimated.

The finite spacial resolution of the microscope used suggests a type A uncertainty, as well, which is estimated to be the size of a single pixel at 40x magnification with 2-by-2 pixel binning \--- $\sigma_\text{A}=\SI{0.323}{\micro\metre\per px}$. The total uncertainty in determining the position of a given edge is a product of $\sigma_\text{A}$ and $\sigma_\text{B}$ and amounts to roughly a micron.

To determine the uncertainty in the thickness, we use Eq. 1, in which $F\pars{\brcs{x_i}}=H\ut{p}\pars{z_1,z_2}=z_1-z_2$ and $\sigma_{z_1}=\sigma_{z_2}=\SI{1}{\micro\metre}$:
\begin{equation}
\sigma_{H_\text{p}}=\sigma_{z_1}\sqrt{2}=\SI{1.4}{\micro\metre}
\end{equation}

The thickness of 20 cylinders, produced in the microfluidic device, is measured. An average thickness $\avg{H\ut{p}}=\SI{25.7}{\micro\metre}$ and sample standard deviation $s=\SI{0.3}{\micro\metre}$ is calculated. The minimum and maximum values are $H\ut{p,min}=\SI[separate-uncertainty=true,multi-part-units=single]{25.0(14)}{\micro\metre}$ and $H\ut{p,max}=\SI[separate-uncertainty=true,multi-part-units=single]{26.1(14)}{\micro\metre}$. Since the uncertainty in the measurements is greater than the sample standard deviation, we report $\avg{H\ut{p}}=\SI[separate-uncertainty=true,multi-part-units=single]{25.7(14)}{\micro\metre}$.

\subsection{Uncertainty in the channel height}

To determine the channel height using optical microscopy, the microfluidic chip is cut with a razor and the cross section is viewed. The type B uncertainty in this case is 2 pixels since the channel edges are sharper than those of PEG-DA particles. The type A uncertainty remains the size of a single pixel. Following the same approach like in the previous case, an uncertainty $\sigma_H=2\times0.323\sqrt{2}=\SI{0.9}{\micro\metre}$ is calculated.

\subsection{Uncertainty in the average flow velocity}

Obtaining the mean flow velocity requires fluorescent particle tracking in which a video of fluorescent beads flowing through the channel is recorded. The positions of a given particle in two consecutive frames are compared and the path length is calculated as their difference. Similarly to the uncertainties in channel and particle height, the uncertainty in the flow velocity scales linearly with the length-to-pixel ratio of the lens used and the uncertainty in determining the center of a given particle. The particle tracking algorithm uses a weighted centroid method to calculate the particle center, which has sub-pixel accuracy. We will make a non-conservative estimate and limit the type B uncertainty to half a pixel.  Then, the uncertainty in the distance traveled by any particle over two consecutive frames is $\sigma_{\Delta x}=0.323/2\sqrt{2}=\SI{0.23}{\micro\meter}$. The frame rate of the movie is 20 FPS, therefore the lag between two consecutive frames is $t_\text{f}=\SI{0.05}{\second}$. Applying Eq. 1 to the expression for the particle velocity $v_\text{p}=\Delta x/t_\text{f}$ yields:
\begin{equation}
\sigma_{v_\text{p}}=\frac{\sigma_{\Delta x}}{t_\text{f}}=\SI{4.6}{\micro\metre\per\second}
\end{equation} 

\subsection{Uncertainty in the disk position in a dumbbell}

The uncertainty in determining where the center of a disk within a dumbbell lies is governed by the spacial resolution of the microscope and the accuracy of the numerical algorithm used to detect the disks. While tracking dumbbells, we use a 20x lens with 2-by-2 pixel binning, which results in a distance-to-pixel ratio of $\SI{0.650}{\micro\metre\per px}$. The circular Hough transform we use to detect the disks utilizes a weighted centroid method, which reports the coordinates of the disk center with a sub-pixel accuracy. However, the algorithm reports radii with a resolution of 1 pixel. Therefore, we estimate an uncertainty of 1 pixel or $\SI{0.650}{\micro\meter}$ in determining the disk coordinates.

\subsection{Uncertainty in the shaft length in a dumbbell}

Although the center-to-center distance of the two disks is fixed, there is both type A and type B uncertainty in its determination. We discussed the type A uncertainty in the previous section. The type B uncertainty stems from the randomness of the photopolymerization reaction \--- different particles have slightly different shaft lengths. To estimate the uncertainty in the length of the shaft, we write the expression for it explicitly and apply Eq. 1 to it:
\begin{equation}
\abs{\vect{O}\pars{t}}=l=\sqrt{O_x^2\pars{t}+O_y^2\pars{t}}=\sqrt{\pars{x\ut{S}\pars{t}-x\ut{B}\pars{t}}^2+\pars{y\ut{S}\pars{t}-y\ut{B}\pars{t}}^2}.
\end{equation}

\begin{equation}
\begin{split}
\sigma_l&=\sqrt{\pars{\pder{l}{x\ut{S}}\sigma_{x\ut{S}}}^2+\pars{\pder{l}{x\ut{B}}\sigma_{x\ut{B}}}^2+\pars{\pder{l}{y\ut{S}}\sigma_{y\ut{S}}}^2+\pars{\pder{l}{y\ut{B}}\sigma_{y\ut{B}}}^2}\\
&=\sqrt{\frac{2\pars{x\ut{S}\pars{t}-x\ut{B}\pars{t}}^2\sigma_{x\ut{S}}^2}{l^2}+\frac{2\pars{y\ut{S}\pars{t}-y\ut{B}\pars{t}}^2\sigma_{y\ut{S}}^2}{l^2}}\\
&=\sqrt{\frac{2\sigma_{x\ut{S}}}{l^2}l^2}=\sigma_{x\ut{S}}\sqrt{2}=\SI{0.9}{\micro\metre}.
\end{split}
\end{equation}

\subsection{Uncertainty in the orientation angle}

The orientation angle of the microparticle $\theta\pars{t}$ is calculated from the vector $\vect{O}\pars{t}$, which starts from the center of the big disk of a dumbbell and ends in the center of the small disk:
\begin{equation}
\theta\pars{t}=\arccos\pars{\frac{\vect{O}\pars{t}\vect{\cdot}\vect{e}_0}{\abs{\vect{O}\pars{t}}\abs{\vect{e}_0}}}=\arccos\pars{\frac{O_x\pars{t}}{\abs{\vect{O}\pars{t}}}}=\arccos\pars{\frac{O_x\pars{t}}{\sqrt{O_x^2\pars{t}+O_y^2\pars{t}}}},
\end{equation} 
where $\vect{e}_0$ is the unit vector aligned with the flow velocity. It is made sure the channel is leveled in such a way that the flow direction is parallel to the $x$-coordinate of the field of view. This reduces the dot product of the two vectors to the $x$-component of $\vect{O}\pars{t}$. Applying Eq. 1 to Eq. 6 yields:
\begin{equation}
\begin{split}
\sigma_\theta&=\sqrt{\pars{\pder{\theta}{x\ut{S}}\sigma_{x\ut{S}}}^2+\pars{\pder{\theta}{x\ut{B}}\sigma_{x\ut{B}}}^2+\pars{\pder{\theta}{y\ut{S}}\sigma_{y\ut{S}}}^2+\pars{\pder{\theta}{y\ut{B}}\sigma_{y\ut{B}}}^2}\\
&=\sqrt{2\frac{\sigma_{x\ut{S}}^2}{l^2}}=\frac{\sigma_{x\ut{B}}}{l}\sqrt{2}=\SI{0.84}{\degree}.
\end{split}
\end{equation}
We make a non-conservative estimate and round the uncertainty in the orientation angle up to $\SI{1}{\degree}$.

\subsection{Uncertainty in the characteristic rotation time $\tau$}

The rotation time is easily accessible numerically via the rotation velocity at $\theta=\pi/2$. Experimentally, however, this is not a viable option because this would mean estimating $\tau$ from only two micrographs with great uncertainty. To obtain a statistically-significant result experimentally, we fit the time evolution of the orientation angle to an analytical expression for $\theta\pars{t}$. Solving Eq. 28 in the main text by making the substitution $\sin\pars{\theta}=2x/\pars{1+x^2}$ yields:
\begin{equation}
-\frac{t}{\tau}=\ln\pars{\frac{\tan\pars{\theta/2}}{\tan\pars{\theta_0/2}}}.
\end{equation}
Rearranging, we obtain the equation in linear form:
\begin{equation}
\ln\pars{\tan\pars{\theta/2}}=\ln\pars{\tan\pars{\theta_0/2}}-\frac{1}{\tau}t.
\end{equation}
We calculate $\tau$ from the slope of a line fitted to a dataset $\left[t,\ln\pars{\tan\pars{\theta/2}}\right]$. The estimated slope has an uncertainty since the data cannot be perfectly fitted to a line. The uncertainty in the slope $b$ is inversely proportional to the square root of the number of data points $N$ used in the linear regression:
\begin{equation}
\sigma_b=b\frac{1}{\sqrt{N-2}}\sqrt{\frac{1}{R^2}-1},
\end{equation}
where $R^2$ reflects the goodness of the fit.
In our experiments the relative uncertainty in the slope is less than $1.5\%$ due to:
\begin{enumerate}
\item the great number of images we record \--- each experimental movie consists of at least 600 frames.
\item the high $R^2$ value \--- in all experiments $R^2>0.980$.
\end{enumerate}

\subsection{Uncertainty in the scaled rotation time $\tilde{\tau}$}
Applying Eq. 1 to the scaled rotation time $\tilde{\tau}=\tau U_0/H$ yields:
\begin{equation}
\begin{split}
\sigma_{\tilde{\tau}}&=\sqrt{\frac{U_0^2}{H^2}\sigma_\tau^2+\frac{\tau^2}{H^2}\sigma_{U_0}^2+\frac{U_0^2\tau^2}{H^4}\sigma_H^2}\\&=\sqrt{\tilde{\tau}^2\frac{\sigma_\tau^2}{\tau^2}+\tilde{\tau}^2\frac{\sigma_{U_0}^2}{U_0^2}+\tilde{\tau}^2\frac{\sigma_H^2}{H^2}}\\&=\tilde{\tau}\sqrt{\frac{\sigma_\tau^2}{\tau^2}+\frac{\sigma_{U_0}^2}{U_0^2}+\frac{\sigma_H^2}{H^2}}
\end{split}
\end{equation}
The three relative uncertainties are $\sigma_\tau/\tau\simeq1\%$, $\sigma_{U_0}/U_0\simeq10\%$ and $\sigma_H/H=3\%$. Filling in these numbers in Eq. 11 yields a relative uncertainty in $\tilde{\tau}$ of roughly $11\%$ \--- the uncertainty is entirely dominated by the uncertainty in $U_0$.

%
\end{document}